\documentstyle[12pt]{article}
\begin{document}
\begin{center}

{\bf Quantum Entangled States and Quasiclassical Dynamics\\
 in Macroscopic Spin Systems}\\ \ \\
Gennady P. Berman$^a$, Gary D. Doolen$^a$, Gustavo V. L\'opez$^b$\\and Vladimir I. Tsifrinovich$^c$\\ \ \\ 
$^a$Theoretical Division and the CNLS, \\
Los Alamos National Laboratory, Los Alamos, New Mexico 87545\\
$^b$ Departamento de F\'isica, Universidad de Guadalajara,\\
 Corregidora 500, S.R. 44420, Guadalajara, Jalisco, M\'exico\\
$^c$Department of Applied Mathematics and Physics, Polytechnic University,\\
Six Metrotech Center, Brooklyn NY 11201
\end{center}
\begin{abstract}
When dealing with macroscopic objects one usually observes quasiclassical phenomena, which can be described in terms of quasiclassical (or classical) equations of motion. Recent development of the theory of quantum computation is based on implementation of the entangled states which do not have a classical analogy. Using a simple example of a paramagnetic spin system we show that the entangled states can be detected in standard macroscopic experiments as a sharp deviation from quasiclassical motion.
\end{abstract}
\newpage
In this paper, we show that quantum entangled states which are used for quantum computation \cite{1}-\cite{5} can be observed, at low temperature, in a macroscopic system as a sharp deviation from a quasiclassical motion. Consider, as an example, a solid which contains a dilute concentration of ions, $A$ and $B$, which have nuclear spins, $I_1=1/2$, $I_2=1/2$, with different gyromagnetic ratios, $\gamma_1$ and $\gamma_2$. (Those also can be electron spins with different $g$-factors, or electron and nuclear spins of the same ion.) In a permanent magnetic field, $H$, these two spins have different resonant frequencies, $\omega_1=\gamma_1H$ and $\omega_2=\gamma_2H$. We assume that the system can be approximately considered as an ensemble of of two-spin molecules, $AB$, and that the spins in each molecule interact via a weak Ising interaction. The Hamiltonian of an individual molecule can be written as,
$$
\hat{\cal H}_0=-\hbar(\omega_1\hat I^z_1+\omega_2\hat I^z_2+2J\hat I_1^z\hat I_2^z),\eqno(1)
$$
where $J$ is the interaction constant. We shall consider the evolution of this system under the action of the electromagnetic pulse with the frequency $\omega$. The Hamiltonian of interaction between a molecule and the rotating magnetic field can be written as,
$$
\hat {\cal H}_1=-{{\hbar}\over{2}}\sum_{i=1}^2\Omega_i\Bigg(e^{-i\omega t}\hat I_i^-+e^{i\omega t}\hat I_i^+\Bigg),\eqno(2)
$$
where $\Omega_1$ and $\Omega_2$ are the Rabi frequencies of spins $A$ and $B$. The wave function of a two-spin molecule can be represented in the form,
$$
\Psi(t)=c_{00}(t)|00>+c_{01}(t)|01>+c_{10}(t)|10>+c_{11}(t)|11>.\eqno(3)
$$
The evolution of the system is described by the Schr\"odinger equation,
$$
i\hbar\dot \Psi=\hat{\cal H}\Psi,\quad (\hat{\cal H}=\hat{\cal H}_0+\hat{\cal H}_1).\eqno(4)
$$
(We assume that the temperature is small in comparison with the Zeeman splitting.)
In the rotating reference frame, we derive the time-independent equations of motion for complex amplitudes, $c_{ik}(t)$, in (3) \cite{6,7}. 

Assume that the system is initially in the state,
$$
\Psi(0)={{1}\over{\sqrt{2}}}\bigg(|0>-i|1>\bigg)|0>.\eqno(5)
$$
In this state, we have the following values of the average spins,
$$
I_1^x(0)=I_1^z(0)=0,\quad I_1^y(0)={{i}\over{2}}(c^*_{10}c_{00}-c_{10}c_{00}^*
+c^*_{11}c_{01}-c_{11}c_{01}^*)=-{{1}\over{2}},\eqno(6)
$$
$$
I_2^x(0)=I^y_2(0)=0,\quad I^z_2(0)={{1}\over{2}}(|c_{00}|^2-|c_{01}|^2+
|c_{10}|^2-|c_{11}|^2)={{1}\over{2}}.
$$
One can see that in the state (5), the left spin (1) points in the negative $y$-direction, and the right spin (2) points in the positive $z$-direction (the direction of the permanent magnetic field). So far, this is a quasiclassical state which corresponds to two spins positing in definite directions. Now, we  apply
a $\pi$-pulse with the frequency  $(\omega_2-J)$ which drives the transition $|10>\rightarrow i|11>$. As a result of the action of this pulse, the system is  transformed into the simplest entangled state,
$$
\Psi={{1}\over{\sqrt{2}}}(|00>+|11>).\eqno(7)
$$
This state does not have a quasiclassical analogy. The average values of both spins are equal to zero in this state,
$$
I^{x,y,z}_1=I^{x,y,z}_2=0.\eqno(8)
$$
In Figs 1 and 2 (on left side, labeled  ``quantum'') we show the time evolution of average spins under the action of a $\pi$-pulse obtained from the equations for the amplitudes $c_{ik}(t)$. One can see in Fig. 1 a monotonic decrease of the amplitude of precession of the left spin (1) from the initial value 0.5 to 0. The value of the $z$-component of the left spin is very small (less than $10^{-3}$) but it also remains near zero at the end of a $\pi$-pulse. 
In Fig. 2, one can see that the $z$-component of the right spin (2) decreases monotonically from 0.5 to 0. The $y$-component reaches its maximum value, 0.25, in the middle of the pulse, and then approaches zero. The $x$-component is very small ($\sim 10^{-3}$), and its amplitude sharply decreases at the end of the $\pi$-pulse. Note, that extremely small oscillations of the $x$-component of the right spin remain after the action of the $\pi$-pulse, indicating that the $\pi$-pulse does not produce a perfect entangled state.

Now we show that the quasiclassical evolution of a two-spin molecule drastically differs from the quantum evolution. The equations of motion for the quasiclassical two-spin molecule can be written as,
$$
\dot{\vec I}_k=-{\vec I}_k\times {{\partial{\cal H}}\over{\partial{\vec I}_k}},\eqno(9)
$$
where the quasiclassical Hamiltonian has the form,
$$
{\cal H}=-(\omega_1I^z_1+\omega_2I^z_2 +2JI^z_1I^z_2).\eqno(10)
$$
The right (indicated as ``classical'') sides in Figs 1 and 2 show the evolution of quasiclassical spins under the action of a $\pi$-pulse, for the initial conditions (6) of the average quantum spins,
$$
I^x_1(0)=I^z_1(0)=0,\quad I_1(0)^y=-{{1}\over{2}},\eqno(11)
$$
$$
I^x_2(0)=I^y_2(0)=0,\quad I_2^z(0)={{1}\over{2}}.
$$
In Fig. 1, one can see a stable precession of the left (1) spin in the $xy$ plane of the rotating frame. In Fig. 2, one can see a very small deviation of the second spin from its initial position (11). So, for the classical two-spin molecule a $\pi$-pulse with the frequency $(\omega_2-J)$ does not influence significantly the dynamics of the spins. This situation can be easily understood. Indeed, for the initial conditions (11), the frequency $(\omega_2-J)$ is not a resonant frequency. In this case, the first spin is in the $xy$ plane, and it does not produce an effective field for the second spin. Thus, the resonant frequency for the second spin is $\omega_2$ rather than  $(\omega_2-J)$.

In conclusion, we have shown that the entangled states can be observed in the experiments with macroscopic systems by indicating a sharp deviation from the quasiclassical behavior. For a considered case of two-spin molecules a $\pi$-pulse with a sertain frequency does not influence significantly a classical dynamics.  But in a quantum case, this $\pi$-pulse  leads to the total vanishing of the longitudional and transversal magnitization of a macroscopic sample. The experiments on creation of entangled states in considered in this paper macroscopic systems require application only one $\pi$-pulse with a proper frequency. Such kind of experiments can significantly improve our understanding of the properties of entangled states, their stability and effects connected with decoherence. 
\section*{ACKNOWLEDGEMENTS}
G.V.L. and V.I.T. are grateful to the Theoretical Division and the CNLS of the Los Alamos National Laboratory for their hospitality.
This work  was partly supported by the Defense Advanced Research Projects
Agency.
\newpage
\begin{center}
{\bf Figure Captions}
\end{center}
\quad\\
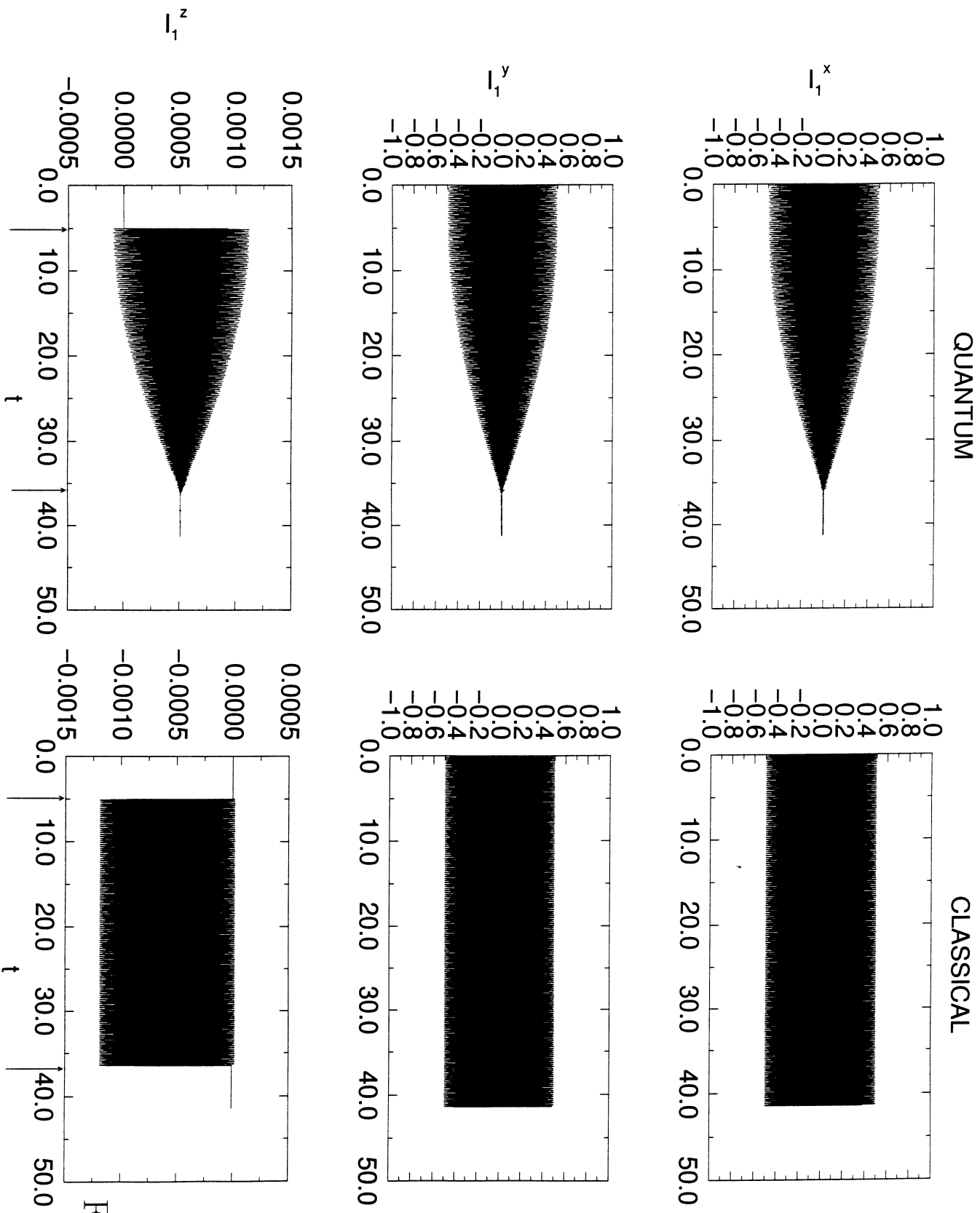
Fig. 1. Time evolution of the $x$, $y$, and $z$ -- components of the average spin, ${\vec I}_1$, (left), and its classical analog (right) under the action of a $\pi$-pulse with the frequency $\omega_2-J$. The values of parameters are: $\Omega_1=5\Omega_2$, $\Omega_1=0.1$, $J=5$. The vertical arrows indicate the beginning and the end of a $\pi$-pulse.
\\ \ \\
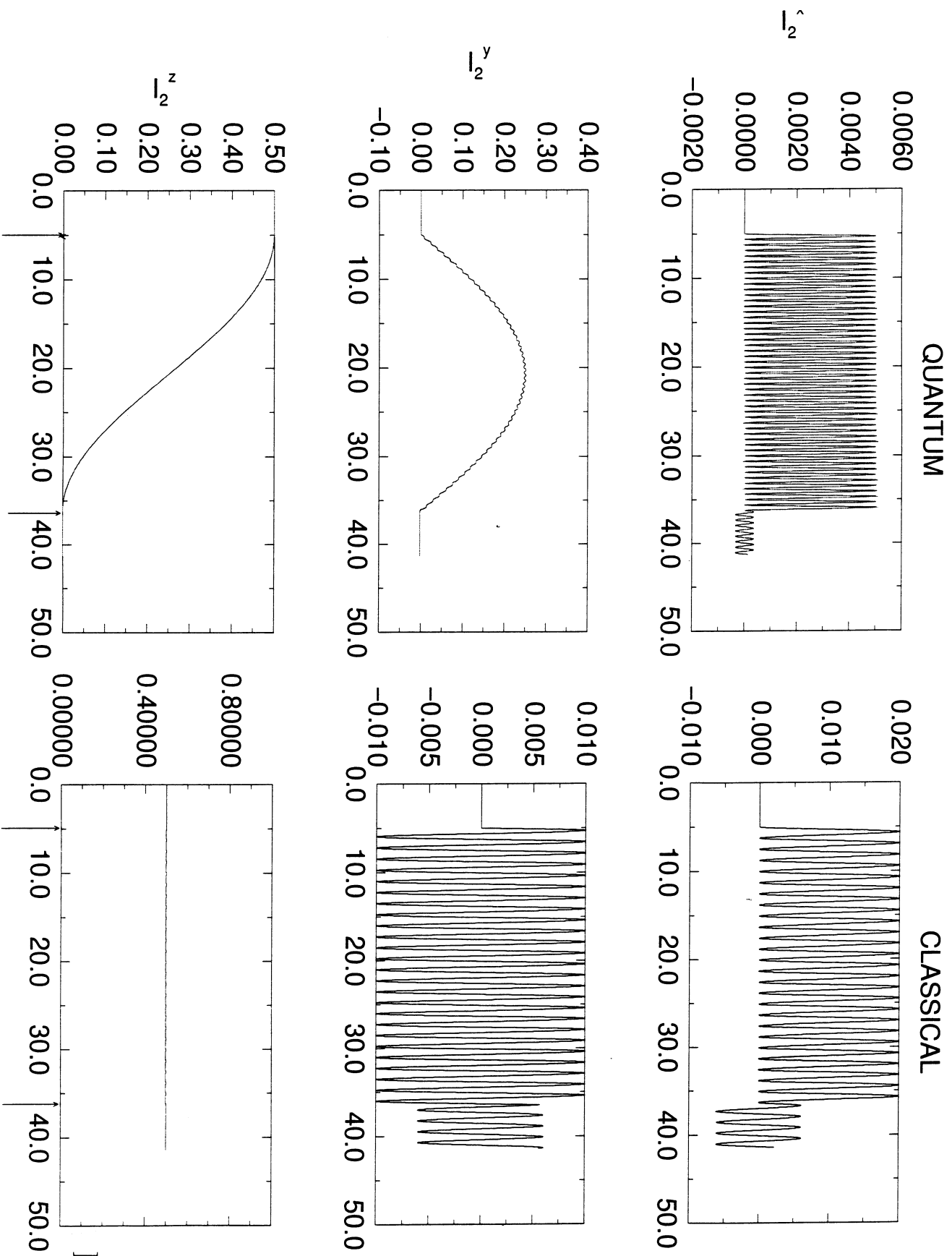
Fig. 2. Time evolution of the $x$, $y$, and $z$ -- components of the average spin, ${\vec I}_2$, (left), and its classical analog (right) under the action of a $\pi$-pulse with the frequency $\omega_2-J$. The values of parameters are the same as in Fig. 1. The vertical arrows indicate the beginning and the end of a $\pi$-pulse.
\newpage
\end{document}